# Switching of the topologically trivial and non-trivial quantum phase transitions in compressed 1T-TiTe$_2$: Experiments and Theory

V. Rajaji[1], Utpal Dutta[2], P. C. Sreeparvathy[3], Saurav Ch. Sarma[4], Y.A. Sorb[1], B. Joseph[5], Subodha Sahoo[2], Sebastian C. Peter[4], V. Kanchana[3], and Chandrabhas Narayana[1,*]

[1]Chemistry and Physics of Materials Unit, Jawaharlal Nehru Centre for Advanced Scientific Research, Jakkur P.O., Bangalore 560 064, India

[2]HP&SRPD, Bhabha Atomic Research Centre, Trombay, Mumbai 400 085, India

[3]Department of Physics, Indian Institute of Technology, Hyderabad, Kandi, Sangareddy 502285, Telangana, India

[4]New Chemistry Unit, Jawaharlal Nehru Centre for Advanced Scientific Research, Jakkur P.O., Bangalore 560 064, India

5 Elettra Sincrotrone Trieste, S.S. 14, Km 163.5 in Area Science Park, Basovizza, Trieste 34012, Italy

**ABSTRACT**

We report the structural, vibrational and electrical transport properties up to ~ 16 GPa of the 1T-TiTe$_2$, a prominent layered 2D system, which is predicted to show a series of topologically trivial - nontrivial transitions under hydrostatic compression. We clearly show signatures of two iso-structural transition at ~ 2 GPa and ~ 4 GPa obtained from the minima in *c/a* ratio concomitant with the phonon linewidth anomalies of E$_g$ and A$_{1g}$ modes at around the same pressures, providing strong indication of unusual electron-phonon coupling associated to these transitions. Resistivity presents nonlinear behavior over similar pressure ranges providing a strong indication of the electronic origin of these pressure driven isostructural transitions. Our data thus provide clear evidences of topological changes at A and L point of the Brillouin zone predicted to be present in the compressed 1T-TiTe$_2$. Between 4 GPa and ~ 8 GPa, the *c/a* ratio shows a plateau suggesting a transformation from an anisotropic 2D layer to a quasi 3D crystal network. First principles calculations suggest that the 2D to quasi 3D evolution without any structural phase transitions is mainly due to the increased interlayer Te-Te interactions (bridging) via the charge density overlap. In addition to the pressure dependent isostructural phase transitions, our data also evidences the occurrence of a first order structural phase transition from the trigonal (*P*$\bar{3}$m1) phase at higher pressures. We estimate the start of this structural phase transition to be ~ 8 GPa and the symmetric of the new high-pressure phase to be monoclinic (*C*2/m).

## I. INTRODUCTION

Recently a new state of quantum matter known as Topological Insulators (TI) have received great attention due to their potential applications in spintronics, quantum computing, and thermoelectric energy conversion devices.[1] TI are a novel class of materials which are insulating in its bulk but supports spin-dependent and time reversal symmetry protected conducting states at the boundaries due to strong spin-orbit coupling (SOC).[1,2] Interestingly, some SOC narrow band gap materials are trivial insulators at ambient conditions but can be transformed into non-trivial topological insulators by applying strain. This transition is named as topological quantum phase transition (TQPT).[3-5] It is an iso-structural second order transition which arises as a consequence of an adiabatic band inversion occurring at the time reversal invariant momenta point (TRIM) with parity change (odd/even). In this process, topological invariant $Z_2$ changes from $Z_2 = 0$ (conventional insulator) to $Z_2 = 1$ (topological insulator).[1,3-5] Generally, strain can be induced into the SOC materials by either chemical or physical routes. For instance, chemical doping in $TiBi(S_{1-x}Se_x)_2$,[6,7] and $Pb_{1-x}Sn_xSe$ systems[8] cause TQPT. Similarly, the experimentally accessible physical strain, i.e., hydrostatic pressure is another ideal external tool to tune the SOC strengths, hybridization, density and crystal field splitting in narrow band gap materials which may induce TQPT. Indeed hydrostatic pressure induced TQPT has been observed in several systems like BiTeI, BiTeBr, and $Sb_2Se_3$.[9-11]

Due to its technological importance, a considerable number of materials have been theoretically predicted as a topological insulator under high-pressure. However, a direct experimental detection of band inversion with the high-pressure setup is challenging to perform. For example, angle resolved photo emission spectroscopy (ARPES) is the most direct tool to probe the non-trivial electronic band inversion.[12,13] But ARPRES measurements under pressure is not yet implemented due to the experimental difficulties. However the indirect evidence of TQPT can be obtained from a combination of transport, synchrotron x-ray diffraction (XRD) and Raman linewidth anomalies which originate from charge density redistribution and electron phonon coupling during TQPT transition.[9-11,14,15] For instance, a combined synchrotron powder XRD and infrared spectroscopy measurements on BiTeI (space group S.G: $P3m1$, band gap $E_g = 0.38$ eV) revealed a correlation between band gap closing and band inversion with a minimum of $c/a$ ratio in the pressure range 2.0 – 2.9 GPa.[9,16] Interestingly, a phonon linewidth anomaly (unusual electron phonon coupling) of E mode at ~ 3.0 GPa has been observed during TQPT in BiTeI.[14] Furthermore, unusual increase in the inner Fermi surface shape and curvature changes of outer Fermi surface shape has been noticed from Shubnikov-de Haas (SdH) oscillations measurements during the TQPT in BiTeI.[17] Similarly, TQPT has been claimed in $Sb_2Se_3$ (SG: $Pnma$, $E_g = 1$ eV) at ~ 2.5 GPa by studying the vibrational phonon and electrical resistivity anomalies together with the first principles calculations,[11,15] though there is also an alternate interpretation suggested.[18] Recently, Ohmura et al., showed that bismuth tellurihalide BiTeBr (S.G: $P3m1$, $E_g = 0.55$ eV) undergoes a TQPT at 2.5 – 3.0 GPa using resistivity and synchrotron

XRD measurements under pressure.[10] The above examples provide strong basis for using such indirect methods to study pressure induced TQPT in the SOC systems.

Titanium based transition metal dichalcogenides (TMD) $TiX_2$ (X = Te, Se, and S) crystallize in layered hexagonal structure (SG: $P\bar{3}m1$, No: 164) which shows exotic properties like charge density wave, superconductivity, etc.[19,20] Among these $TiTe_2$ has recently been received significant interest due to their series of topological transitions under moderate pressures and potential usage for information processing.[21,22] The unit cell of 1T-$TiTe_2$ consists of stacks of hexagonal close packed layers of Ti metal atom sandwiched between two adjacent layers of Te atoms and in each layer, Ti atom is octahedrally surrounded by six Te atoms. It has predominately weak Van der Waals-type interlayer bonding forces along the c axis and strong intralayer covalent bonds along ab plane. The first principles calculations based on density functional theory (DFT) predicted a series of pressure induced transitions between topologically trivial and non-trivial phases related to the band inversions at different points (L, M and Γ) of the Brillouin zone in $TiTe_2$.[22] This remarkable theoretical prediction strongly motivated us to explore the pressure induced topological changes in $TiTe_2$ compound through XRD, Raman scattering, and electrical transport measurements. To the best of our knowledge, till date, there have been no experimental studies reported on the 1T-$TiTe_2$ under pressure.

In this paper, we present the structural, vibrational and electrical transport properties of 1T-$TiTe_2$ under hydrostatic compression for the first time. The synchrotron XRD, Raman scattering, and electrical transport anomalies show signatures of the two iso-structural electronic transitions at ~ 2 GPa and ~ 4 GPa in the 1T phase, which we have attributed to the non-trivial TQPT and the trivial metallic transition, respectively, based on the recent theoretical report.[22] Further, the applied pressure switches the 2D layered material (anisotropic) into isotropic 3D crystal above ~ 4 GPa through charge density overlapping between the interlayer Te atoms along the c axis. The experimental evidence of isotropic 3D behavior (constant $c/a$ ratio) was explained using the first principles theoretical calculations. This is followed by the 1T phase undergoing a pressure induced structural transition from trigonal (S.G: $P\bar{3}m1$) to monoclinic (S.G: $C2/m$) phase at ~ 8 GPa.

## II. EXPERIMENTAL DETAILS

The 1T phase of $TiTe_2$ was synthesized by mixing 0.1579 g of titanium shots (99.99 %, Alfa Aesar) and 0.8420 g of tellurium shots (99.99 %, Alfa Aesar) in a 9 mm diameter quartz tube. The tube was flame-sealed under the vacuum of $10^{-3}$ Torr, achieved with the help of a rotary pump, to prevent oxidation during heating. The tube was then placed in a vertically aligned tube furnace and heated to 800 °C over a period of 8 h to allow proper homogenization. Subsequently, the temperature was kept constant for 6 days. Finally, the system was allowed to cool to room temperature over a period of 10 h. No reaction with the quartz tube was observed. A black polycrystalline $TiTe_2$ was formed.

Raman spectra were recorded using WITec micro Raman spectrometer (UHTS600) in the backscattering geometry (180˚). The Raman spectrometer equipped with a diode pumped frequency doubled Nd:YAG solid state laser (wavelength λ = 532 nm), 600 mm focal length mono-chromator and Peltier air cooled CCD detector. The spectral resolution is about ~ 0.5 cm$^{-1}$ for the grating of 2400 lines per mm. The *in situ* high-pressure Raman scattering measurements were performed using a membrane type diamond anvil cell (DAC) with the culet size of 400 μm. A T301 stainless steel gasket with the starting thickness of about ~ 250 μm was pre-indented to the thickness of about ~ 60 μm. Then a hole of ~ 150 μm diameter was drilled at the center which acts as the sample chamber, and the pressure was calculated by ruby fluorescence method.[23] A mixture of methanol:ethanol (4:1) was used as the pressure transmitting medium (PTM) which guarantees the hydrostatic limit up to ~ 10 GPa and quasi hydrostatic limit up to 25 GPa.[24] The accumulation time of each spectrum was about 5 minutes. The lower value of laser power (< 0.5 mW) was maintained to avoid the risk of heating and oxidation of the samples.

The *in situ* high-pressure synchrotron XRD experiments were carried out using a Mao-Bell type DAC with diamonds having a culet size of 400 μm. The synchrotron radiation XRD measurements were performed at the XPRESS beamline of Elettra, Treste, Italy using the monochromatic radiation with energy of E = 24.762 KeV (λ = 0.50070 Å). The procedures of gasket preparation, PTM and pressure calibration is the same as mentioned above. The XRD patterns were collected using MAR345 image plate detector. Typical exposure time was about 4 minutes for each pattern. The calibration of a sample to detector distance and the image plate orientation angles were carried out using LaB$_6$ as standard. The two dimensional (2D) XRD image patterns were converted into the one dimensional (1D) intensity versus diffraction angle (2θ) patterns using the Fit2D software.[25]

Pressure dependent electrical resistance was measured up to ~ 16 GPa at room temperature by standard quasi-four probe method using a miniature DAC and an ac-resistance bridge in combination with fine gold electrodes fabricated on the diamond culet. The sample and electrodes were insulated from the metal gasket using an insulation layer of Al$_2$O$_3$ and epoxy mixture. The sample pressure was measured by *in situ* ruby fluorescence method at any temperature.[23] Powdered NaCl was used as the PTM which not only maintains quasi-hydrostaticity but is also used to keep the electrodes in good contact with the sample.

### III. COMPUTATIONAL METHOD

The calculations were carried out within the frame work of density functional theory (DFT) implemented in CASTEP and WIEN2k packages.[26,27] The experimental parameters are considered as an input, and the structure is optimized using Broyden- Fletcher -Goldfarb- Shanno (BFGS) minimization scheme.[28] The optimized structure has been used to calculate the bonding and electronic structure properties, which was performed using WIEN2k package with generalized gradient approximation of Perdew, Burke, and Ernzerhof (GGA-PBE) functional.[29] Considering the presence of heavy elements we have included spin orbit coupling in the calculations. A dense k mesh of 39× 39×19 was used and all the calculations were performed with the optimized lattice parameters with an energy convergence criterion of $10^{-6}$ Ry per formula unit. Raman spectra were calculated with CASTEP package.

## IV. RESULTS

### A. Characterization of TiTe$_2$ at ambient condition

The Rietveld refinement of the XRD pattern for the $P\bar{3}m1$ structure (referred to as 1T phase) as shown in the Fig. 1(a). The calculated cell parameters and volume at ~ 0.36 GPa are $a$ = 3.76416 Å, $c$ = 6.46711 Å and $V$ = 79.355 Å$^3$ respectively, which show good agreement with the 1T phase of the previous report at ambient conditions.[30,31] The typical unit cell for the 1T structure as shown in the Fig. 1(b). There are three atoms in the unit cell of the 1T-TiTe$_2$, where Ti$^{4+}$ and Te$^{2-}$ atoms occupy 1a and 2d Wyckoff sites, respectively. Further, the presence of a small elemental Te has been detected in synchrotron pattern and is indicated by green color asterisk symbol in Fig. 1(a). We have carefully excluded the Te regions during the refinements.

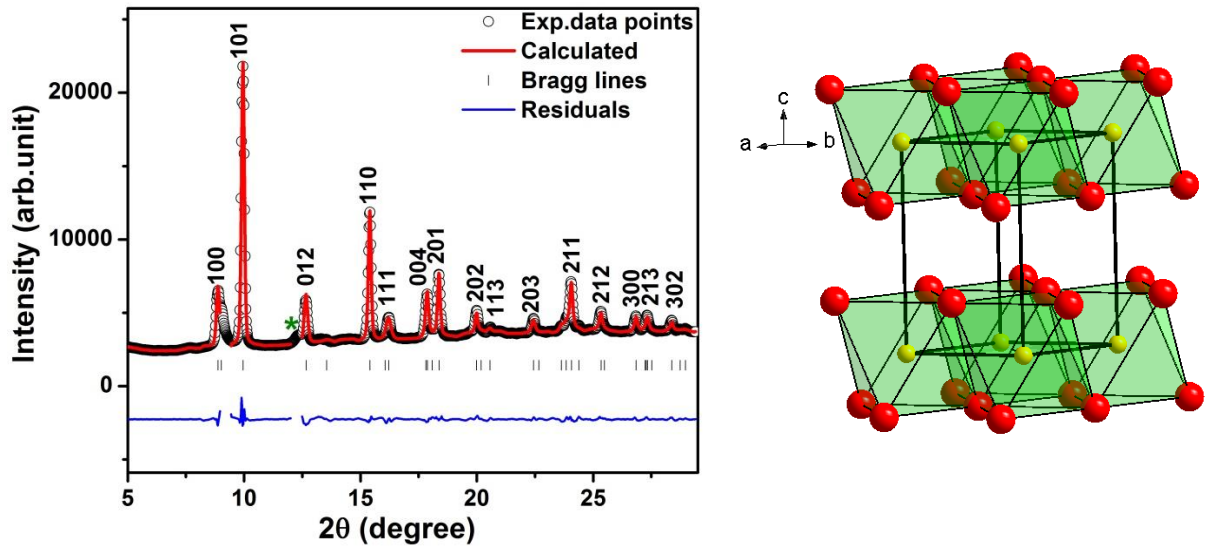

**FIG. 1** (a) Rietveld refinement of the synchrotron XRD pattern of 1T-TiTe$_2$ at ~ 0.36 GPa and (b) schematics of the unit cell of 1T-TiTe$_2$. The yellow and red color atoms represent the Ti and Te, respectively.

According to group theoretical analysis, the layered 1T structure of TiTe$_2$ has nine vibrational modes at the gamma point of the phonon dispersion curve.[32,33]

$$\Gamma = E_g + A_{1g} + 2E_u + 2A_{2u}$$

where, the *gerade* ($E_g$ and $A_{1g}$) and *ungerade* ($E_u$ and $A_{2u}$) modes represent the Raman active and IR active phonon modes, respectively. In this centro-symmetric structure, doubly degenerate $E_g$ mode (symmetric in plane bending) represents the atomic vibrations along the ab plane whereas $A_{1g}$ mode (symmetric out of plane stretching) represents the atomic vibrations parallel to c axis as shown in the Fig. 2.

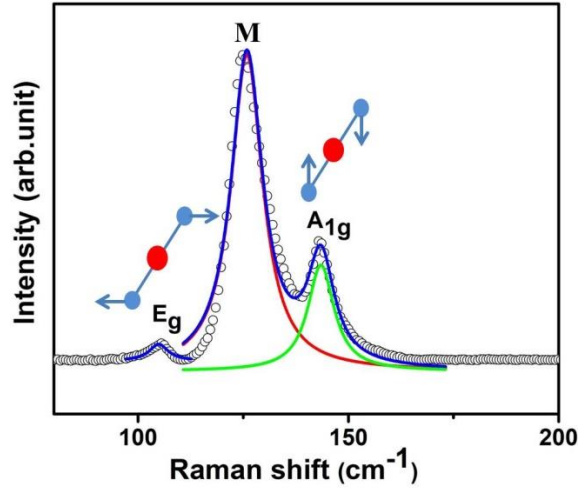

**FIG. 2.** Raman spectrum of 1T-TiTe$_2$ at ambient conditions.

**TABLE I.** The assignment of the Raman modes for 1T-TiTe$_2$.

| Raman mode | Experimental frequency (cm$^{-1}$) | | | Theoretical frequency (cm$^{-1}$) | | |
|---|---|---|---|---|---|---|
| | This work | Ref.34[a] | Ref.21[b] | This work | Ref.33 | Ref.32 |
| $E_g$ | 105 | 102 | - | 100 | 105 | 99.1 |
| M | 126 | - | 124 | - | - | - |
| $A_{1g}$ | 143 | 145 | 145 | 140 | 150 | 145.1 |

[a] single crystal, [b] few layers

Raman modes were fitted using Lorentzian line shape function. Based on our theoretical calculation and the existing literature, the phonon modes at ~ 105 cm$^{-1}$ and ~ 143 cm$^{-1}$ are assigned to $E_g$ and $A_{1g}$ symmetry respectively.[32-34] However, we observed an additional strong mode at ~ 126 cm$^{-1}$, named as M mode. This mode was seen in a few layers of TiTe$_2$ grown as a thin film by Khan et al., and was attributed to $E_g$ symmetry.[21] This assignment seems to be unreliable since the polarization dependent study on single crystal confirmed that selection rule allowed two Raman active modes ($E_g$ and $A_{1g}$) and its energies are ~ 102 cm$^{-1}$ and ~ 145 cm$^{-1}$, respectively.[34] Recent accurately calculated vibrational modes of TiTe$_2$ closely match with our assignment.[32] Hence this M mode could be a zone-folded Raman active mode and has been observed in proto type 1T phase layered TMD materials at different conditions.[35-37] However, the polarization and temperature dependent behavior of this mode (M) in the few layers may give more insight into this mode, which is the subject of future interest. The detailed comparative analysis of vibrational energies for the 1T-$TiTe_2$ compound is shown in Table I.

## B. Synchrotron XRD measurements under pressure

*In situ* high-pressure synchrotron XRD measurements were carried out up to ~ 16 GPa and the representative XRD plots for selected pressures are shown in Fig. 3. The systematic increase in the Bragg peaks to higher diffraction angle (2θ) is consistent with the compression of the unit cell. Furthermore, the appearances of new Bragg peaks at ~ 12.0 GPa indicates structural transition. However, the onset of phase transition point can be traced to ~ 8 GPa via the (101) and (110) Bragg peaks analysis (by peak fitting) and this has been commented in the supplementary materials (Figs. S1(a) and S1(b)). By comparing our XRD patterns (Fig.3) with proto type compounds like $IrTe_2$ and $ZrS_2$ ($CdI_2$ type structure), we found that $TiTe_2$ follows an identical structural sequence with $IrTe_2$ and $ZrS_2$ under high-pressure.[38,39] Notably, a distinct splitting of (101) Bragg peak observed in $TiTe_2$ is exactly in agreement with $IrTe_2$ and $ZrS_2$.[38,39] Even though a new high-pressure phase appears in the $TiTe_2$, the ambient phase coexists up to ~ 16.0 GPa, the maximum pressure reached in this study. The structural evolution of Te phase (shown as green asterisk in Fig. 3) under pressure is well established, hence the discussion on the high-pressure phases of Te were excluded in the whole pressure range of this study.

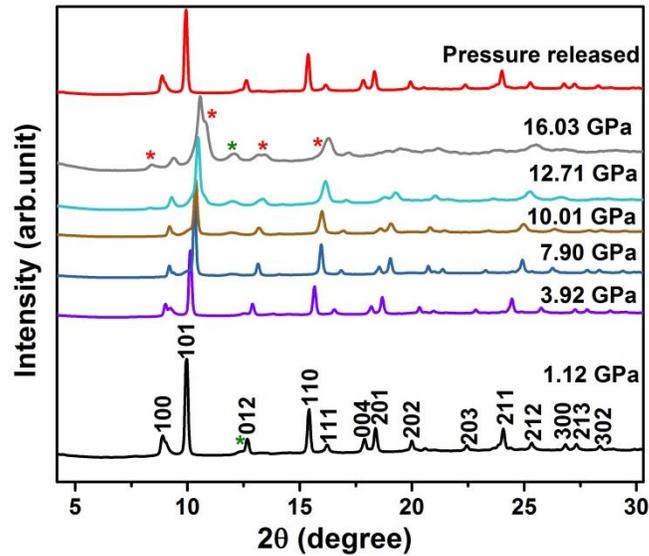

**FIG. 3.** Pressure evolution of the synchrotron XRD patterns of $TiTe_2$ at selected pressure values. The red color asterisk symbol represents the appearance of new Bragg peaks at higher pressure regions.

In the 1T phase, the only free atomic position is z coordinate of the Te ions which defines the Te(1) – Te(2) contact distance. The lattice parameters and atomic coordinates are refined using the FullProf software[40] for each XRD pattern up to 8 GPa. After ~ 8 GPa, due to the complexity of the mixed phases, we have analyzed the 1T phase of TiTe$_2$ using Powd and Dicvol software,[41] which provides only the unit cell parameters (a, b, c) and volume (V). Figure 4 represents the systematic decrease in volume of the unit cell up to ~ 16 GPa. Notably, an apparent change is observed at ~ 8 GPa, which further ascertain the phase transition. The pressure-volume data best fit the equation of state (EOS) into two different regions using the following Murnaghan EOS and third order Birch Murnaghan EOS respectively.[42,43]

$$P(V) = \frac{B_0}{B_0'}\left[\left(\frac{V_0}{V}\right)^{B_0'} - 1\right]$$

$$P(V) = \frac{3B_0}{2}\left[\left(\frac{V_0}{V}\right)^{\frac{7}{3}} - \left(\frac{V_0}{V}\right)^{\frac{5}{3}}\right]\left\{1 + \frac{3}{4}(B_0' - 4)\left[\left(\frac{V_0}{V}\right)^{\frac{2}{3}} - 1\right]\right\}$$

where, $B_0$, $B_0'$ and $V_0$ are the isothermal bulk modulus, the derivative of bulk modulus and volume at room pressure, respectively. The Murnaghan EOS was used to fit the pressure region up to ~ 8 GPa and the fit yields $V_0$ = 80.34 Å$^3$, bulk modulus $B_0$ = 28.60 GPa and $B_0'$ = 7.19. In the mixed phase regions (8 – 16 GPa), the 1T phase was fitted by third order Birch Murnaghan EOS, and the fit gives $V_0$ = 79.26 Å$^3$, bulk modulus $B_0$ = 40.73 GPa and $B_0'$ = 6.02. After the phase transition, $B_0$ increases from 28.60 GPa to 40.73 GPa suggesting that the high-pressure 1T phase has lesser compressibility than ambient conditions.

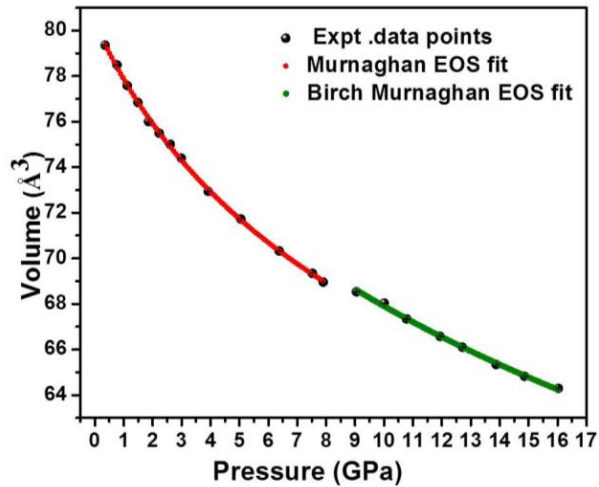

**FIG. 4.** EOS fit for the 1T-TiTe$_2$ phase to the pressure versus volume data.

The pressure dependence of the normalized lattice parameters ($a/a_0$, $c/c_0$) is plotted in Fig. S2. Though $a/a_0$ and $c/c_0$ decreases systematically under pressure up to ~ 16 GPa, a clear anomaly in $c/c_0$ at ~ 8 GPa is observed (see supplementary material). Fig. 5 represents the pressure versus *c/a* ratio of 1T-TiTe$_2$. Initially, the *c/a* ratio decreases from 1.718 to 1.690, implies that *c* axis is more compressible than *a* axis which is usually expected for anisotropic layered crystals due to the weak van der Waals interlayer forces along *c* direction. Interestingly, two inflection points are noticed in 1T phase at ~ 2 GPa and ~ 4 GPa. Here, we would like to emphasize that a similar trend was reported in the pressure range 2.0 – 2.9 GPa and 2.5 – 3.0 GPa in BiTeI and BiTeBr, respectively and these changes were interpreted as the signature of TQPT.[9,10] To get more insight about *c/a* ratio anomalies, the pressure dependent Te(1) – Te(2) contact distance is plotted in Fig. S3. As seen from Fig. S3, it shows two distinct anomalies in the 1T phase region at ~ 2 GPa and ~ 4 GPa. But after 4 GPa, *c/a* ratio surprisingly turns out to be almost pressure invariant which suggests that the compressibility of both the lattice parameters (*a* and *c*) are similar. The plausible cause for this behavior could be due to the lower threshold level of Te(1) – Te(2) anionic contact distance is reached and strong charge repulsion (coulomb) built up between the interlayers. Hence this constant behavior of *c/a* ratio under pressure hints the isotropic nature. So, the pressure switches the 2D layered 1T-TiTe$_2$ into quasi 3D network like feature from 4 GPa to 8 GPa and similar observation has been made in MoSe$_2$.[44] This change in axial compressibility is directly related to the fluctuations in the charge density distribution along the different directions, as we discuss in more detail below (theoretical results). During the pressure regions 4 – 8 GPa, the huge amount of strain is developed inside the sample. In order to relax the strain, the 1T phase undergoes a structural phase transition. Evidently, the discontinuity in *c/a* ratio at ~ 8 GPa indicating the structural transition and further it increases with pressure. Therefore, in 1T phase, initially *a* axis is stiffer than *c* axis and after the structural transition, *c* axis is stiffer than *a* axis.

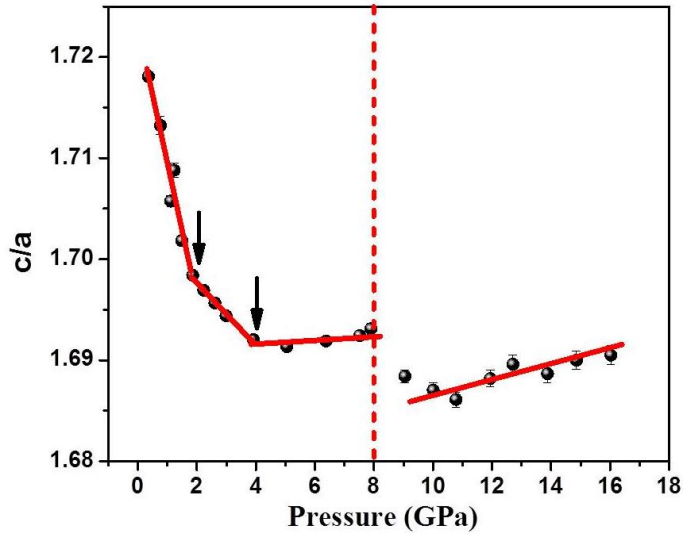

**FIG. 5.** Pressure dependence of the *c/a* ratio for 1T-TiTe$_2$. The solid and vertical dotted lines signify guide to the eye and structural phase transition respectively.

The IrTe$_2$ also undergoes pressure induced structural transformation from $P\bar{3}m1$ to $C2/m$ at ~ 5 GPa.[38] Due to the poor data quality and complexity of the mixed phase, we could not refine this phase through Rietveld method from ~ 8 GPa onwards. However, we have indexed the XRD pattern at ~ 13.90 GPa with monoclinic $C2/m$ space group using the Powd and Dicvol software[41] as shown in the Fig. S4. The indexed values for the monoclinic unit cell at ~ 13.90 GPa are $a$ = 17.3666 Å, $b$ = 3.5545 Å, $c$ = 5.6966 Å, $\beta$ = 91.17°, $V$ = 351.57 Å$^3$, Z = 6 and V/Z = 58.60 Å$^3$, which agree well with the similar proto type system, IrTe$_2$.[38] The volume change of ~ 9.5 % (when we extrapolate the volume data of the high-pressure phase to ~ 8 GPa) is observed during the structural transition which confirms the first order nature of the transition (see Fig. S5). Mention must be made that V/Z values of monoclinic phase ($C2/m$), which very well agrees with the V/Z trend of $P\bar{3}m1$ phase as shown in the Fig. S3. During the indexing of monoclinic $C2/m$ phase, the "$b$" axis is considered as the unique axis, and the lattice parameters of the indexed pattern for three different pressure values are given in the Table SI (see Supplementary materials). Upon releasing pressure, the high-pressure phase transformed back to the ambient 1T phase indicating reversibility of the transition. The study of detailed pressure induced structural changes with atomic coordinates is beyond the scope of the present work, which will be the future interest of our group.

**C. Raman scattering measurements under pressure**

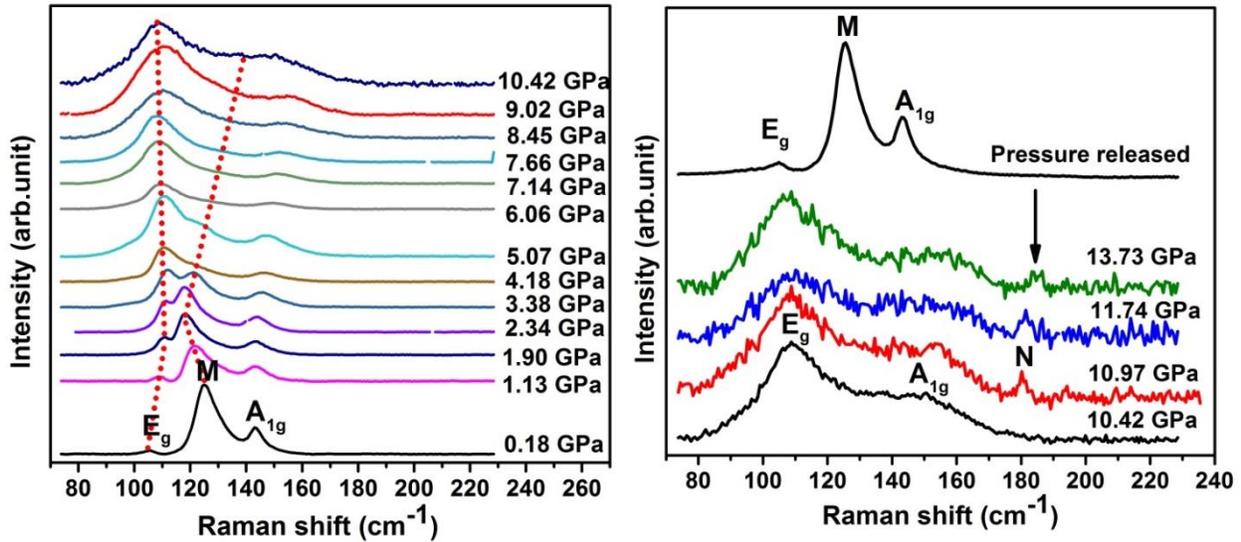

**FIG. 6.** (a) The representative Raman spectra of TiTe$_2$ at relatively low pressure regions and (b) at high-pressure regions and depressurized ambient Raman spectrum.

To shed light on the observed isostructural anomalies ($c/a$ ratio) and phase transitions in TiTe$_2$, Raman spectroscopy measurement under pressure was employed up to ~ 13.7 GPa. The pressure evolutions of Raman spectra of TiTe$_2$ are shown in the Figs. 6(a) and 6(b). As the pressure increases, the intensity of E$_g$ mode increases, whereas the intensity of M and A$_{1g}$ modes are decreasing. However, the overall intensity of all the phonon modes are observed to be drastically decreased above ~ 8.0 GPa. As evident from Fig. 6(b), the appearance of a new Raman mode at ~ 10.97 GPa (named as N mode) confirms the structural phase transition and the presence of $E_g$ and $A_{1g}$ modes at higher pressures confirms the phase coexistence (mixed phase), which is consistent with the XRD results. After ~ 13.7 GPa, the peaks become very broad and difficult to deconvolute it from the background. During the depressurization, the system came back to initial phase (1T-TiTe$_2$), which suggests the observed transition is reversible.

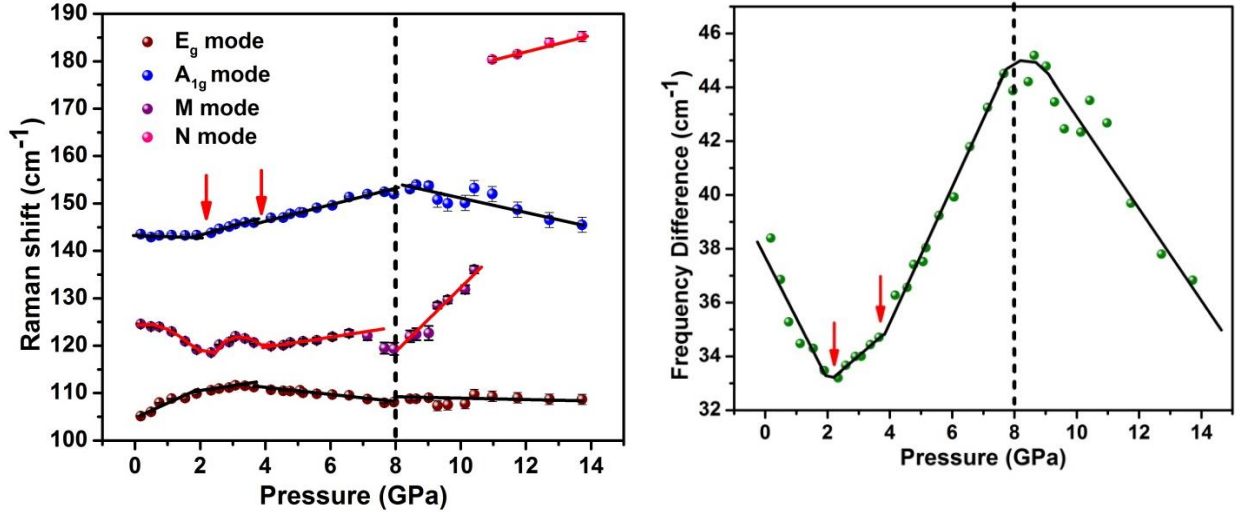

**FIG. 7.** (a) Pressure versus Raman shift of phonon modes ($A_{1g}$, M and $E_g$) of TiTe$_2$. The solid black line represents the linear fit, and the red line represents guide to the eye. (b) Pressure versus frequency difference ($A_{1g}$-$E_g$) between the $A_{1g}$ and $E_g$ modes of TiTe$_2$. The solid red arrows at ~ 2 GPa and ~ 4 GPa represent the isostructural electronic transitions. The solid and vertical dotted lines represent guide to the eye and structural phase transition respectively.

The pressure dependent Raman shift of $E_g$, $A_{1g}$, M and N modes are shown in the Fig. 7(a). In general, the phonon modes are expected to harden (blue shift) during the hydrostatic lattice compression. But, Fig. 7(a) shows the mode behaviors of all the modes are anomalous and we can identify four distinct pressure regions. To elucidate this we have fitted in each region $A_{1g}$ and $E_g$ modes using linear equation and the fitting parameters (slope $a_1$ and intercept $\omega(P_0)$) are summarized in Table II. The $A_{1g}$ mode softening slightly up to ~ 2 GPa and thereafter it starts to harden up to ~ 8 GPa with a small change in slope at ~ 4 GPa. While, the $E_g$ mode shows hardening up to 4 GPa with a clear change in slope at ~ 2 GPa followed by softening till 8 GPa. Upon further compression, the frequency of $E_g$ mode and the newly appeared N mode starts to increase, while the $A_{1g}$ mode begins to soften up to ~ 14 GPa, which is the maximum pressure achieved in Raman study. The frequency of zone folded Raman mode M exhibits very interesting high-pressure behavior. It shows two parabolic pressure dependences with two distinct points of inflections at ~ 2 and ~ 4 GPa, beyond which it slowly hardens up to ~ 8 GPa. After the phase transition, M mode shows significant hardening with pressure. Notably, over the pressure range between 0 – 2 GPa and also between 4 – 8 GPa, the pressure dependence of $A_{1g}$ and $E_g$ modes show opposite behaviors.

**TABLE II.** Pressure dependence behavior of various Raman-mode frequencies and Gruneisen parameters ($\gamma$) of 1T-TiTe$_2$. The pressure coefficients for 1T-TiTe$_2$ were fitted[46] using $\omega(P) = \omega(P_0) + a_1 \cdot (P - P_0)$. The Gruneisen parameters $\gamma$ are determined by using the relation $\gamma = (\frac{B}{\omega(P0)} \times \frac{\partial \omega}{\partial P})$, where B represents the bulk modulus.

| Raman Mode | $\omega(P_0)$ (cm$^{-1}$) | $a_1$ (cm$^{-1}$ GPa$^{-1}$) | $\gamma$ |
|---|---|---|---|
| $E_g$ | 105.1 ± 0.6 [a] | 3.07 ± 0.67 [a] | 0.84 |
|  | 109.3 ± 0.7 [b] | 0.65 ± 0.24 [b] | 0.17 |
|  | 113.6 ± 0.4 [c] | -0.68 ± 0.07 [c] | -0.17 |
|  | 106.5 ± 2.2 [d] | 0.20 ± 0.02 [d] | 0.08 |
| $A_{1g}$ | 143.4 ± 0.1 [a] | -0.10 ± 0.01 [a] | -0.03 |
|  | 140.2 ± 0.6 [b] | 1.69 ± 0.20 [b] | 0.34 |
|  | 139.5 ± 0.6 [c] | 1.70 ± 0.10 [c] | 0.35 |
|  | 165.9 ± 2.9 [d] | -1.46 ± 0.27 [d] | -0.36 |

[a] Estimated at room pressure (P$_0$ = 1atm), [b] Estimated at P$_0$ = 1.89 GPa,
[c] Estimated at P$_0$ = 4.1 GPa, [d] Estimated at P$_0$ = 7.95 GPa.

The drastic softening of $A_{1g}$ mode ($a_1 = -0.68$ cm$^{-1}$/GPa) and hardening of M mode at ~ 8 GPa hint the structural instability and plausible reason for the impending structural phase transition. The slope change of $A_{1g}$ and $E_g$ modes at ~ 8.0 GPa is attributed to the onset of structural phase transition from trigonal ($P\bar{3}m1$) to monoclinic ($C2/m$). The intensity and linewidth of N mode is smaller compared to that of $A_{1g}$ and $E_g$ modes. This could be the reason we did not observe the appearance of N Mode at the onset pressure (~ 8 GPa) of the structural transition. However, once its intensity evolves under pressure, it comes out above 10.97 GPa. To get more insight, the frequency difference between the $A_{1g}$ and $E_g$ modes are plotted as a function of pressure and represented in the Fig. 7(b). The plot illustrates four different regions, which substantiate the analysis of pressure dependence of $A_{1g}$ and $E_g$ modes. The maxima in frequency difference ($A_{1g} - E_g$) at ~ 8 GPa represents the structural phase transition, whereas the two minima at ~ 2GPa and ~ 4 GPa are signifying the isostructural anomalies.

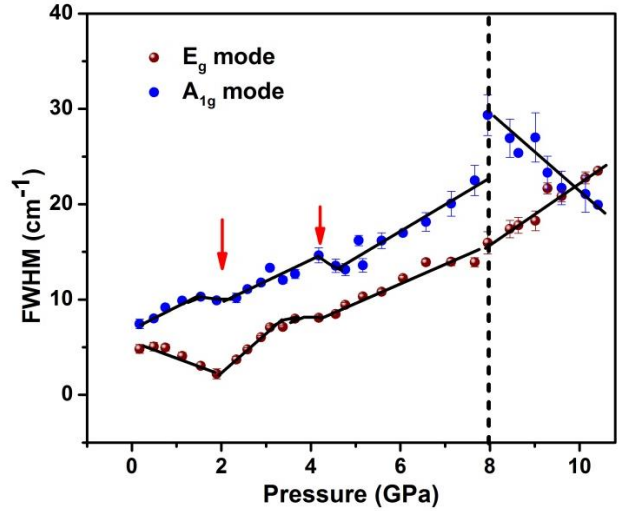

**FIG. 8.** Pressure dependence of FWHM of $A_{1g}$ and $E_g$ modes for $TiTe_2$. The solid red arrows at ~ 2 GPa, and ~ 4 GPa indicate the isostructural electronic transitions. The solid and vertical dotted line represent guide to the eye and structural phase transition respectively.

Raman linewidth studies could provide the information about the phonon-phonon interactions and the excitation-phonon interactions such as electron-phonon, spin-phonon coupling existing in the system.[45-49] Therefore, we have carefully analyzed the FWHM of $A_{1g}$ and $E_g$ modes and are shown in the Fig. 8. It should be noted that the nature of PTM limits the accuracy of information about intrinsic linewidth of sample beyond the hydrostatic limit. Since the methanol-ethanol (4:1) mixture gives only the hydrostatic pressure limit up to ~ 10.5 GPa, therefore the linewidth of phonon modes has been analyzed below 10.5 GPa.[23] Generally, for the crystal, Raman linewidth is inversely proportional to the life time of the phonon modes. It is normally seen that as we increase the pressure, we observe an increase in linewidth of phonon modes. However, the FWHM of $E_g$ mode decreases under pressure up to ~ 2 GPa, followed by an increase up to ~ 10.5 GPa with anomalous behaviors at ~ 4 GPa and ~ 8 GPa. It is noteworthy that the similar pressure induced decrease in linewidth of E mode and $E_g$ modes were observed in BiTeI and $A_2B_3$ (A=Bi, Sb and B= Te, Se and S) series compounds during the TQPT at 3 – 4 GPa and electronic topological transitions (ETT) at 3 – 4 GPa respectively.[14,46,49] In contrast, the $A_{1g}$ linewidth increases up to 8 GPa with significant anomalies at ~ 2 GPa and ~ 4 GPa, followed by a decrease up to 10.5 GPa with a discontinuity observed during the phase transition at ~ 8.0 GPa. After the structural transition, the decreasing trend in linewidth of $A_{1g}$ phonon mode could be due to decrease in electron-phonon coupling in the monoclinic $C2/m$ phase. The zone folded Raman mode M shows increasing linewidth behavior under pressure up to ~ 11 GPa, which is the expected behavior for any phonon (See Fig. S6). More importantly, evidence of an unusual electron-phonon coupling from the linewidth anomalies ($A_{1g}$ and $E_g$) at ~ 2 GPa and ~ 4 GPa further confirms the isostructural ($P\bar{3}m1$) transitions, which could possibly be originated from electronic state modulation under pressure. Finally, pressure dependent the frequency and the linewidth behavior of both $A_{1g}$ and $E_g$ modes suggests two isostructural transition and a structural transition, which is well consistent with the XRD measurement.

## D. Electrical transport measurements under pressure

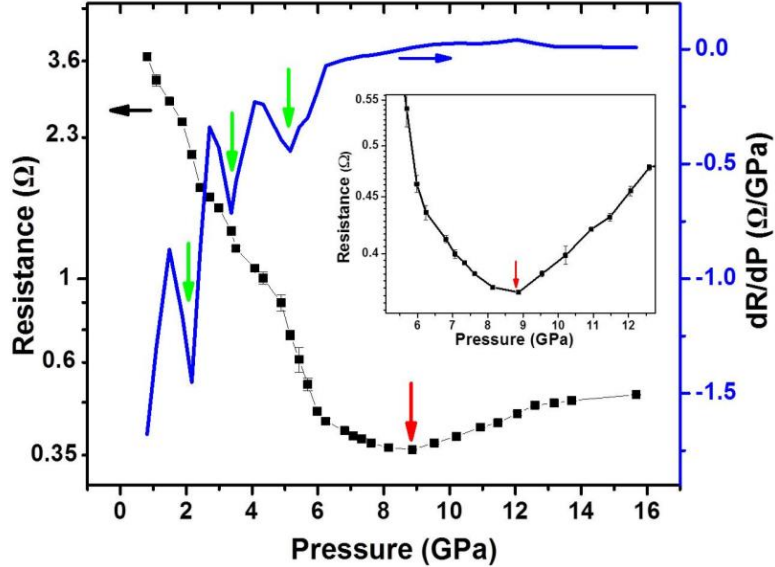

**FIG. 9.** Pressure dependent electrical resistance of TiTe$_2$ (black square and a line corresponding to the left y axis) and its first derivative (blue line corresponding to the right y axis). The red arrow at ~ 8.8 GPa indicates the structural transition. The solid green arrows at 2.1 GPa and 3.4 GPa indicate the isostructural electronic transitions in the trigonal ($P\bar{3}m1$) phase. The inset highlights the resistance minimum in the vicinity of the structural transition.

1T-TiTe$_2$ is expected to show metallic behavior due to the finite band overlap of d orbitals of Ti atom with p orbitals of Te atoms at ambient conditions.[50] The oxidation state of typical CdI$_2$ structures like TiX$_2$ is given by Ti$^{4+}$(X$^{2-}$)$_2$ [X=S, Se, and Te], here the amount of electron transfer from orbitals p to d is zero.[51] However the overlap of the p orbitals with d orbitals can lead to the transfer of n electrons per metal, then the oxidation of formula can be changed into Ti$^{(4-n)+}$(X$^{(2-(n/2))-}$)$_2$.[51] In the energy band diagram, the transition metal Ti d orbitals are located just above the top of p orbitals of Te chacolgen.[51] These two orbitals can be overlapped either via the chemical or physical methods. Chemically, it can be achieved by decreasing the electronegativity of chalcogen X. As the electronegativity of Te is less than both Se and S, the top portions of the p orbital bands are raised. Hence, the overlap of p-d bands is more in Te than Se and S atoms which lead to the behavior of TiTe$_2$, TiSe$_2$ and TiS$_2$ as metal, semimetal and semiconductor, respectively. Physically, the overlap of the p-d orbitals can be increased by reducing the Ti-X bond distance, which can be experimentally achieved using hydrostatic pressure.

The pressure dependence of the electrical resistance (R) and its first derivative (dR/dP) at room temperature for 1T-TiTe$_2$ are illustrated in Fig. 9. As the pressure increases, the resistance of TiTe$_2$ sample quickly drops from ~ 3.7 Ω at ~ 0.8 GPa to ~ 0.37 Ω at ~ 8.8 GPa. As we further increase the pressure from 8.8 GPa the resistance starts to increase slowly with pressure (clearly seen in the inset of the Fig. 9) and at ~ 12.6 GPa reaches a value of ~ 0.48 Ω which is 30 % more than that at ~ 8.8 GPa. Above ~ 12.6 GPa the resistance increases at a slower rate and the 0.51 Ω resistance at ~ 16 GPa (the highest measured pressure of our experiment) is roughly 38 % more than at ~ 8.8 GPa value. The increase in resistance may be caused by sample size shrinkage. Fritsch et al.[52] suggested that the increase in resistance by the sample size shrinkage is about one third of the compressibility, which in the present case should be less than 10 % below ~ 8.8 GPa and 13 % at 16 GPa based on the bulk modulus measurements of our XRD experiment. Thus, our result suggests that, as the pressure increases, TiTe$_2$ becomes more and more metallic only up to 8.8 GPa and the unusual increase of the resistance above ~ 12.6 GPa should mainly result from the accompanying change in the crystal structure above 8.8 GPa which is consistent with the structural transition from trigonal ($P\bar{3}$m1) to monoclinic ($C$2/m)) as confirmed by XRD and Raman measurements at ~ 8 GPa. This type of change in crystal symmetry along with abnormal resistance increase with pressure was also observed in V$_2$O$_3$.[53] It is also reported that the pressure-induced structural phase transitions of Bi$_2$Te$_3$ and As$_2$Te$_3$ induce a series of changes in the electrical resistivity.[54,55]

In the low-pressure regime (below 8.8 GPa), pressure dependent resistance curve shows three distinct slope changes at ~ 2.1 GPa, ~ 3.4 GPa and ~ 5.1 GPa which are identified by the minima of the P vs. dR/dP curve. These inflection points cannot be associated with structural phase transitions since high-pressure XRD and Raman measurements reveal the structural stability of the ambient-pressure $P\bar{3}$m1 structure up to ~ 8 GPa and are associated with isostructural electronic transitions. The first two points (~ 2.1 GPa and ~ 3.4 GPa) are consistent with our XRD and Raman measurements (~ 2 GPa and ~ 4 GPa) and could be due to the TQPT. But the third transition point at ~ 5.1 GPa is not seen in XRD and Raman measurement. We have seen from XRD that the strains build up in the pressure range of 4 − 8 GPa followed by a structural phase transition at ~ 8.8 GPa. In addition, we observe the broad nature of the minimum at ~ 5.1 GPa of the dR/dP curve. Hence, the anomaly at ~ 5.1 GPa may be the signature of the precursor effect for the structural transition. We observe a considerable hysteresis between the pressure increasing, and the pressure decreasing cycle, which confirms the first order nature of the transition at ~ 8.8 GPa (see Fig. S7). The isostructural and structural transition observed in resistance studies is consistent with the XRD and Raman measurements and the small difference in pressure value could be mainly due to the sensitiveness of these technique, error in pressure measurement and the degree of hydrostatic conditions produced by PTM used in these experiments. More importantly, the pressure dependent electrical resistance measurement confirms the isostructural transitions at ~ 2.1 GPa and ~ 3.4 GPa which is in electronic in origin.

**E. FIRST PRINCIPLES CALCULATIONS**

First principles studies here were carried out to analyze the quasi 3D nature of 1T-TiTe$_2$ under compression via the charge density redistribution. Here, we have analyzed the inter layer Te(1) – Te(2) bonding, intra layer Ti – Te bonding and charge density plots. Bond lengths of both inter layer Te(1) – Te(2) and intra layer Ti – Te are plotted as a function of pressure as shown in the Fig. 10. Due to the layered nature, the inter layer Te(1) – Te(2) bond length is higher than the intra layer Te-Ti bond length at ambient condition. Under the application of pressure above 2 GPa the intra layer Ti – Te bond length is more than the inter layer Te – Te bonding, which might result in the reduction of 2D character and increased overlap of orbitals between the p and d.

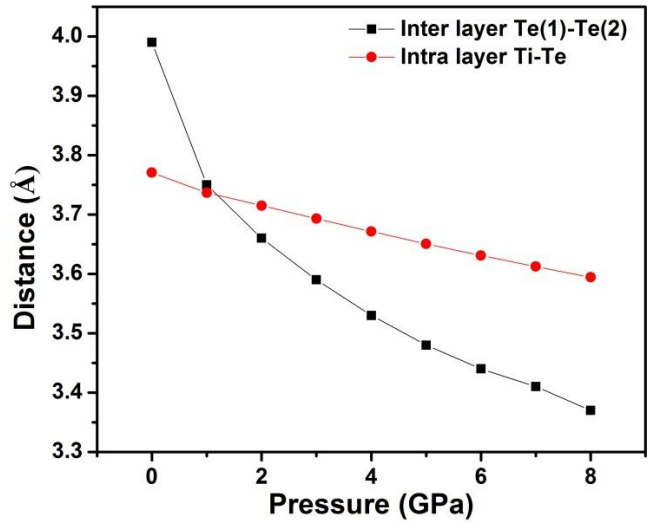

**FIG. 10.** Pressure dependence of the interlayer Te(1)-Te(2) and intra layer Ti-Te distance.

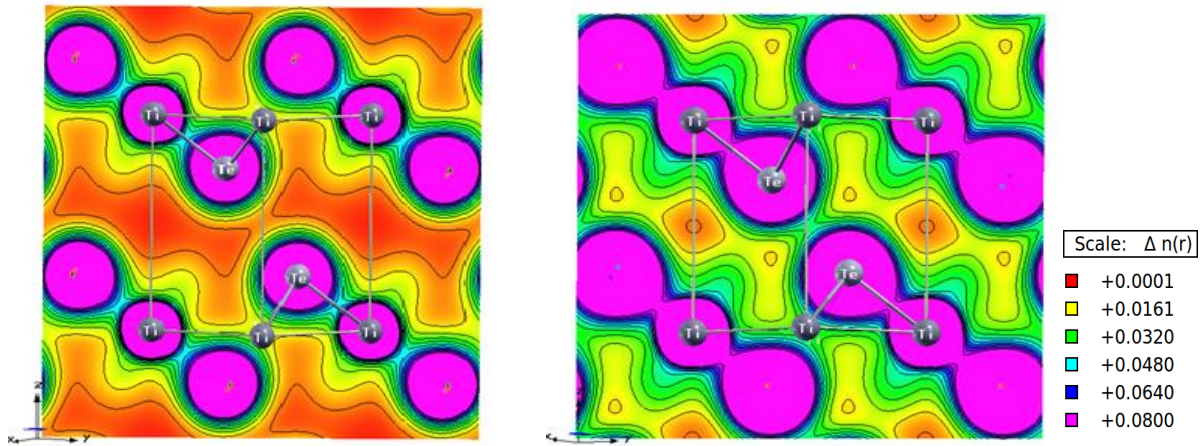

**FIG. 11.** (a) Pressure dependent charge density redistribution of (111) plane at 0 GPa and (b) 8 GPa. The relative scale of charge density is given in color code.

Likewise, the bond length between Te(1) and Te(2) decreases with compression, which will increase the charge flow between these two layers and is clearly seen in charge density plots (Fig. 11). Figures 11(a) and (b) represent the charge density plots ((111) plane) both at ambient and 8 GPa respectively (for other pressure values, see Fig. S8). At ambient, we observer a more ionic nature between the inter layer Te(1) and Te(2). With pressure, this ionic nature is found to decrease, and covalent nature is found to increase. In the Fig. 11, we have shown the intra layer Ti – Te bonding as a function of pressure and we observe a large overlap between intra layer Ti-Te, which will cause more charge flow between them. In addition to this, the layer thickness is found to decrease with pressure, and the Te-Ti-Te angle is found to increase with pressure.

## V. DISCUSSION

The recent theoretical calculation showed that 1T-TiTe$_2$ undergoes a series of topological transitions under hydrostatic isothermal compression.[22] 1T-TiTe$_2$ is shown to have four consecutive band inversions at A, L, Γ and A points corresponding to the theoretical pressure points of 3, 8, 15 and ~ 26 GPa respectively.[22] Due to the first band inversions at A point (3 GPa), the system possesses non-trivial TQPT due to the changes in the parity and consequently the topological invariant changed to $Z_2 = 1$. This is followed by another band inversion (~ 8 GPa), which takes place at L point of the Brillouin zone. This leads the system to become a trivial metallic phase due to the net parity change is same with respect to ambient condition ($Z_2 = 0$).[22] Furthermore, there is another band inversions (~ 15 GPa) at Γ point, which changes the overall parity and hence the topological invariant ($Z_2$) changes from 0 to 1, leading to a second non-trivial TQPT. Finally, the band inversion (~ 26 GPa) at A point in the BZ makes the system switches back to trivial metallic phase ($Z_2 = 0$). Interestingly, it is suggested that if there are no phase transitions this cycle of multiple oscillations of topological transition should continue.[22]

The experimentally observed multiple isostructural (1T phase) electronic transition signatures are closely consistent with the above proposed model. We attribute the isostructural anomalies at ~ 2 GPa from XRD, Raman, and resistance to the non-trivial TQPT as consequence of band inversion at A point of the BZ.[22] In this electronic transition, conduction band characters (dominated by Te-p orbitals) are exchanged with valence band characters (dominated by Ti-d orbitals) at A points of the electronic band structure.[22] Similarly, the anomalies at ~ 4 GPa are due to the trivial metal as a consequence of band inversion at L point of the BZ. Here, conduction band characters (dominated by Ti-d orbitals) are switched with valence band characters (dominated by Te-p orbitals) at L points of the electronic band structure.[22] The charge density fluctuations have occurred during band inversion at A and L points of the BZ leads to anomalies in *c/a* ratio at ~ 2 GPa and ~ 4 GPa, respectively. The charge redistribution modulates the electronic structure and consequently the phonon life time is affected, that reflect as the unusual electron-phonon coupling at ~ 2 GPa and ~ 4 GPa in Raman linewidth. A difference exists between the theoretical (3 GPa and 8 GPa) and experimental (2 GPa and 4 GPa) pressure values. It could be mainly due to the approximation used in the theoretical calculations.

The DFT based first principle theoretical calculations support the stability of structural symmetry ($P\bar{3}m1$) of the 1T-TiTe$_2$ up to 30 GPa (hydrostatic pressure).[22,33] But, our experimental results contradict these theoretical proposals and shows the structural transformation of 1T-TiTe$_2$ from trigonal ($P\bar{3}m1$) to monoclinic ($C2/m$) at ~ 8 GPa under hydrostatic pressure. But it should be noted that the trigonal phase persists up to 16 GPa in our experiments, this probably could explain why the theoretical calculations are unable to see the structural transition. This pressure induced structural transition limits the detection of theoretically predicted other two electronic transitions at higher pressure regions ~ 15 GPa and ~ 26 GPa in the $P\bar{3}m1$ phase. Moreover, as the pressure increases, the TiTe$_2$ sample becomes more and more metallic, which is consists with the overall intensity reduction of the phonon modes in Raman measurement. This pressure enhanced metallization could mainly come from the following two physical reasons. (1) The pressure decreases the interlayer Te(1)-Te(2) contact distance, van der Waals interactions and hence bridges the two layers at ~ 4 GPa onwards. (2) In the intralayer, electron transfer from p orbitals of Te atom to d orbitals of Ti atom increases under pressure. This is confirmed by our first principles calculations, which shows that the applied hydrostatic pressure bridges the interlayer Te(1) and Te(2) via the charge density redistributions which results in the conversion of an anisotropic 2D to isotropic 3D behavior at the pressure range from 4 GPa to 8 GPa (1T phase).

## V. CONCLUSIONS

In conclusion, the systematic pressure dependent synchrotron XRD, Raman and electrical resistance studies were carried out on 1T-TiTe$_2$ sample up to ~ 16 GPa. We observe a first order structural phase transition at ~ 8 GPa from trigonal ($P\bar{3}m1$) to monoclinic ($C2/m$) symmetry. The pressure dependent *c/a* ratio and electrical resistance show anomalies at ~ 2 GPa and ~ 4 GPa in the 1T phase which suggests charge density fluctuations upon compression. This is consistent with the phonon linewidth anomalies at ~ 2 GPa and ~ 4 GPa indicating unusual electron-phonon coupling arising from the electronic structure changes under pressure. These multiple experimental signatures of the two isostructural electronic transitions at ~ 2 GPa and ~ 4 GPa are closely consistent with the theoretical predictions and are attributed to non-trivial TQPT and trivial metallic transition, respectively. The 2D layered crystal of TiTe$_2$ (at ambient condition) switched into a quasi 3D network above 4 GPa via shortening of the inter layer Te(1) – Te(2) contact distances by external hydrostatic pressure inducing strains, which could be the precursor for the structural transition observed. We hope our experimental finding will stimulate researchers to further explore this 1T-TiTe$_2$ compound on the aspect of quantum oscillations measurement like Shubnikov de Haas effect under pressure.

**ACKNOWLEDGEMENTS**

We acknowledge Department of Science and Technology (DST), India for supporting financially to carry out XRD measurements at the Xpress beam line, Elettra synchrotron, Trieste, Italy. CN would like to acknowledge Dr. S. Karmakar of BARC, Mumbai for helping us with the high pressure resistance measurements and helping in scientific discussions. CN thanks the Sheikh Saqr Laboratory for the Senior Fellowship.pressure resistance measurements and helping in scientific discussions. CN thanks the Sheikh Saqr Laboratory for the Senior Fellowship.